\documentclass[twocolumn]{article}

\usepackage{fullpage}
\usepackage{epsfig}
\usepackage{url}

\begin{document}
\title{Analyzing Worms and Network Traffic using Compression}
\author{
Stephanie Wehner\\
\emph{Centrum voor Wiskunde en Informatica}\\
\emph{Kruislaan 413, 1098 SJ Amsterdam, Netherlands}\\
wehner@cwi.nl
}

\maketitle

\begin{abstract}
Internet worms have become a widespread threat to system and network operations.
In order to fight them more efficiently, it is necessary to analyze newly discovered worms 
and attack patterns.
This paper shows how techniques based on Kolmogorov Complexity can help in the analysis
of internet worms and network traffic. Using compression, different species of
worms can be clustered by type. This allows us to determine whether an unknown
worm binary could in fact be a later version of an existing worm in an extremely simple,
automated, manner. This may become a useful tool in the initial analysis of malicious
binaries. Furthermore, compression can also be useful to distinguish different types of
network traffic and can thus help to detect traffic anomalies: Certain anomalies
may be detected by looking at the compressibility of a network session alone.
We furthermore 
show how to use compression to detect malicious
network sessions that are very similar to known intrusion attempts. This technique
could become a useful tool to detect new variations of an attack and thus help to
prevent IDS evasion.
We provide two new plugins for Snort which demonstrate both approaches.
\end{abstract}

\section{Introduction}

Recent years have seen a sharp increase in internet worms, such as Sasser and
Msblaster, causing damage to millions of systems worldwide.
Worms are automated programs that
exploit vulnerabilities of computers connected to the network in order to take control of them. Once they have
successfully infected a system, they continue to search for new victims. When a worm finds
a new vulnerable computer it will infect it as well. Worms thus spread through the network on their own.

Given the enormous problems caused by worms, it is important to develop defenses against them.
For this we need to analyze them in detail. Once a system is found to be infected, we would like to
know how this happened, what actions the worm might perform and how we can prevent infection in the future. This 
paper is concerned with one of the 
\emph{first steps} in such a diagnosis: we want to determine, whether we have seen this worm before or whether 
it is very similar to a known worm. To achieve this, we analyze the binary executable of the worm.
Many known worms, such as Sasser, often come in a number of variants. These are usually successive 
versions released by the same author to extend the worm's functionality.  All such versions
together then form a family of related worms. Using compression, we first
cluster the different versions of worms by family. 
We compare an unknown worm to a number of known worms using compression, which often allows 
us to guess its family which simplifies further analysis. This very simple approach does
not make use of any manual comparisons of text or especially selected patterns contained in the 
worm used in conventional analysis. Many binaries found in the wild are also compressed using UPX, 
which is then modified to prevent decompression. This makes it very difficult to analyze
the binary using approaches which rely on disassembling the binary and comparing the flow 
of execution of two programs. Surprisingly, our approach still 
works even if UPX is used, although it becomes less accurate.
Therefore compression may be a useful tool in the initial analysis of newly captured worms. 

Infected systems and network attacks often cause variations in network traffic. In order to recognize
and prevent such attacks one would like to detect these variations. Typically, anomalies 
are detected by searching for specific patterns within the network traffic. For example, Snort~\cite{snort}
can trigger an alarm when the string ``msblast.exe'' is found back in a TCP session. This works
very well, once it is known what patterns we need to look for. Here, we are interested in recognizing
anomalies \emph{even if we don't know what to look for}. First of all, 
we explore the complexity of different 
types of network traffic using compression. The differences in complexity allow us to make a 
rough guess what class of application protocols gave rise to a certain network interaction.
Intuitively, the complexity of an object is determined by how well it compresses. If something compresses well we 
say it has low complexity. If, on the other hand, it can hardly be compressed at all, we say it has high complexity.
For example, the complexity of SSH sessions is very high. If we observe TCP sessions on the same port
with much lower complexity, we suspect an anomaly. This requires no additional knowledge of the anomaly
or a search for specific traffic patterns.
The compress Snort plugin, allows Snort to trigger
an alarm if the complexity of the observed traffic does not fall within a specified range.

Finally, we make use of compression to compare two different sessions. An attacker or a new version 
of a worm may generate slightly different network traffic, which no longer contains the specific pattern we were
looking for and thus escape detection. Here, we are interested in recognizing patterns which are very 
similar to an existing ones. Informally, we say that two protocol sessions are 
close together if their 
combined complexity is not much larger than their individual complexity. This is the case if two sessions 
are very similar and thus describe each other very well. We can use this to determine whether an 
observed TCP session is similar to a prerecorded session corresponding to a known attack. 
The NCD Snort plugin allows Snort to trigger an alarm if an observed session is at a certain distance
from a given session. Even though compression is a relatively expensive operation to perform on network traffic,
we believe this approach could be useful to detect new types of attacks. Once such attacks have been recognized,
specific patterns can be constructed and used in conventional intrusion detection.

\subsection{Related Work}

Graph-based methods have been used with great success in order to compare executable objects by Halvar Flake~\cite{halvar:malware}
as well as Carrera and Erd\'elyi~\cite{carrera:malware}. Recently, Halvar Flake has also been applied this to 
the analysis of malware~\cite{halvar:personal}. Using these methods it is possible
to gain information about the actual security problems used by worms, by comparing different versions of the same
executable. This requires disassembly of the binary. Whereas this approach has the advantage of disclosing
actual information about the nature of the security vulnerability, our focus here is very different. We provide
a fast method for guessing the family of an observed worm without the need for disassembly, which can be 
used in he initial analysis of the executable.
Based on this analysis, other methods can be applied more accurately. 

Much work has been done to determine the nature of network traffic. What makes our approach fundamentally 
different is that we do not make any preselection of patterns or specific characteristics. We leave it entirely
to the compressor to figure out any similarities. 

Kulkarni and Bush~\cite{kulkarni:management} have previously considered methods based on Kolmogorov complexity 
to monitor network traffic. Their approach however is different as they do not use compression 
to estimate Kolmogorov complexity. Furthermore they argue that anomalies in network traffic stand out
in that their complexity is lower than 'good' network traffic. Whereas we find this to be true for 
network problems caused by for example faulty hardware prompting a lot of retransmissions, we argue
that anomalies are generally characterized by a \emph{different} complexity than we would normally 
expect for a certain protocol. For example, if normally we expect only HTTP traffic on port 80 outgoing
and we now find sessions which have a much higher complexity, this may equally well be caused by a malicious
intruder who is now running an encrypted session to the outside. In this specific example, the complexity
has increased for malicious traffic.

Evans and Barnett~\cite{evans} compare the complexity of legal FTP traffic to the complexity of attacks
against FTP servers. To achieve this they analyzed the headers of legal and illegal FTP traffic. For this
they gathered several hundred bytes of good and bad traffic and compressed it using compress.
Our approach differs in that we use the entire packet or even entire TCP sessions. We use this as we believe 
that in the real world, it is hard to collect several hundred bytes of bad traffic from a single attack 
session using headers alone. Attacks exploiting vulnerabilities in a server are often very short and will not 
cause any other malicious traffic on the same port. This is especially the case in 
non-interactive protocols such as HTTP where all interactions consist of a request and reply only. 

Kulkarni, Evans and Barnett~\cite{kulkarni:ddos} also try to track down denial of service attacks using
Kolmogorov complexity. They now estimate the Kolmogorov complexity by computing an estimate of the
entropy of 1's contained in the packet. They then track the complexity over time using the method
of a complexity differential. For this they sample certain packets from a single flow and then compute 
the complexity differential once. Here, we always use compression and do not aim to detect DDOS attacks.

\subsection{Contribution}

\begin{itemize}
\item We propose compression as a new tool in the initial analysis of internet worms.
\item We show that differences in the compression ratio of network sessions can help to distinguish
different types of traffic. This leads to a simple form of anomaly detection which does not assume
prior knowledge of their exact nature.
\item Finally, we demonstrate that compression can help to discover novel attacks which are very similar
to existing ones.
\end{itemize}

\subsection{Outline}

In Section~\ref{ncd}, we take a brief look at the concept of normalized compression distance, which we
will use throughout this paper. Section~\ref{worms} then shows how we can apply this concept to the analysis
of internet worms. In Section~\ref{traffic} we examine how compression can be useful in the analysis of internet
traffic. Finally, in Section~\ref{snort} and Section~\ref{snort2} we use these observations to construct two snort plugins to help 
recognize network anomalies.

\section{Preliminaries}\label{ncd}

In this paper we use the notion of \emph{normalized compression distance} (NCD)
based on Kolmogorov 
complexity~\cite{paul}. Informally, the Kolmogorov complexity $C(x)$ of a finite binary string $x$ is equal to the 
length of the shortest program which outputs $x$. 
The Kolmogorov complexity is 
in fact machine independent up to an additive constant. This means that it does not matter which
type of programming language is used, as long as one such language is fixed.
This is all we will need here. More information can be found in the book by Li and Vitanyi~\cite{paul}.

What does this have to do with compression? We can regard the compressed version of a string $x$ as a program to 
generate $x$ running on our ``decompressor machine''. Thus the Kolmogorov complexity with respect to this machine
cannot be any larger than the length the compressed string. 
We thus estimate $C(x)$ by compression: We take a standard compressor (here bzip2 or zlib) and compress $x$ 
which gives us $x_{compressed}$. We then use $length(x_{compressed})$ in place of $C(x)$. In our case, the string $x$
will be the executable of a worm or the traffic data.

The normalized compression distance (NCD) of two string $x$ and $y$ is given by
$$
\mbox{NCD}(x,y) = \frac{C(xy) - min\{C(x),C(y)\}}{max\{C(x),C(y)\}}.
$$
More details about the NCD can be found in~\cite{rudi}. The  NCD is a very intuitive notion: 
If two strings $A$ and $B$ are very similar, one ``says'' a lot about the other. Thus given string $A$,
we can compress a string $B$ very well, as we only need to encode the differences between $A$
and $B$. Then the compressed length of string $A$ alone will be not very different
from the compressed length of strings $A$ and $B$ together. However, if $A$ and $B$ are rather different, 
the compressed length of the strings together will be significantly larger than the compressed length of 
one of the strings by itself.
For clustering a set of objects we compute the pairwise NCD of all objects in the set. This gives us a distance
matrix for all objects. We then fit a tree to this matrix using the quartet method described in~\cite{rudi}.
Objects which appear close to each other in the tree, are close to each other in terms of the NCD.

To construct the trees displayed throughout this paper we make use of the CompLearn toolkit~\cite{complearn}. 
For information about networking protocols we refer to~\cite{tbaum}.
Information about the different Worms we analyzed can be found in the Virus Encyclopedia~\cite{virus}.

\section{Analyzing Worms}\label{worms}

We now apply the notion of NCD to analyze the binary executables of internet worms.
Once we identified a malicious program on our computer we try to determine
its relation to existing worms. Many worms exist in a variety of different
versions. Often the author of a worm
later makes improvements to his creation thereby introducing new versions.
This was also in the case of the recent Sasser worm.
We call such a collection of different versions a family.
The question we are asking now is:
Given a worm, are there more like it and which family does it belong to?

For our tests, we collected about
790 different worms of various kinds: IRC worms, vb worms, linux worms and
executable windows worms. Roughly 290 of those worms are available in more
than one version. 

\subsection{Clustering Worms}

First of all, we examine the relations among different families of worms using clustering.
Our method works by computing the NCD between all worms used in our experiment. This gives us 
a distance matrix to which a tree is fit for visualization. In this section, we use bzip2 as our compressor.

\subsubsection{Different classes of Worms}
To gain some insight into the relations among several families of worms, we first cluster worms
of completely different types. For this we make use of text based worms (such as IRC worms, labeled mIRC and IRC)
and worms whose names contain PIF targeted at windows systems. Worms prefixed with
Linux are targeted at the Linux operating system. All other worms are aimed at Windows systems and are
binary.

\begin{figure*}[p]
\begin{center}
\epsfig{file=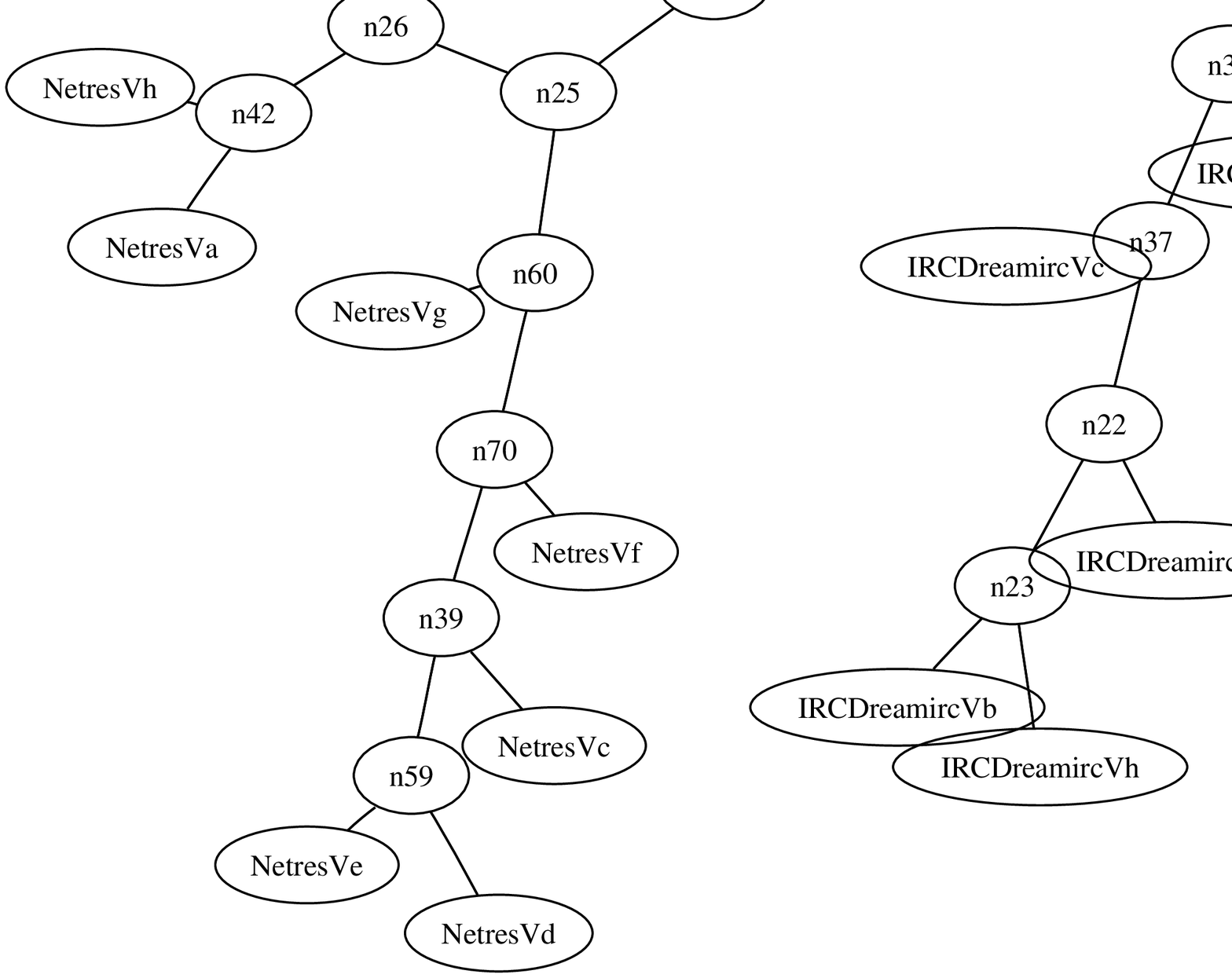, width=18cm}
\caption{Different Classes of Worms}
\label{collection}
\end{center}
\end{figure*}

Looking at Figure~\ref{collection} we see that text based worms have been clustered well together.
We also see a split between binary and text based worms as expected. This in itself is of course not
very interesting. Nevertheless, this naive application of our method yielded some interesting surprises
which caused us to completely reexamine our worm data. 
What is the LinuxGodog worm among the text based worms? Upon closer inspection we discovered that unlike the other
Linux worms which were all binary executables, LinuxGodog was a shell script, thus being too similar
to the script based IRC worms. A similar discovery was made when looking at the two windows worms
Fozer and Netsp, which are both windows batch scripts. We also found out that SpthJsgVa was mislabeled,
and is an IRC worm very closely related to the IRCSpth series. In fact, manual inspection revealed that
it must be another variant. Similarly, IRCDreamircVi was clustered with the other irc worms. It appeared
that this was the only non-binary member of the IRCDreamirc series. 
Another interesting fact was revealed by clustering the Alcaul worm.
A large number of different Alcaul worms were available, all in binary format. However during this
test we accidentally picked the only one which was mime encoded and thus appears closer to text than to
the binary worms. Our test also puts binary Linux worms closer together then binary worms for other
systems. Only LinuxAdm appears further
apart. We attribute this to the fact that binaries of linux and windows are still not too far apart
even if they are of different executable formats. Worms of the same family are put together. We can for example see
a clear separation between the IRCSpth and the IRCSimpsalapim series of text based worms. 
However, the separation of the Mimail and Netres family shows that this approach is also applicable to
binaries.
The MyDoom worm was initially thought to be an exception. The two versions of MyDoom we obtained, 
were not as similar as other worms of equal families.
Closer examination, however, revealed that the MyDoom.a we used was in fact a truncated version of
the binary. 
In general, our initial overview clusters the different worm types well. 

\subsubsection{Windows Worms}

\begin{figure}[p]
\begin{center}
\epsfig{file=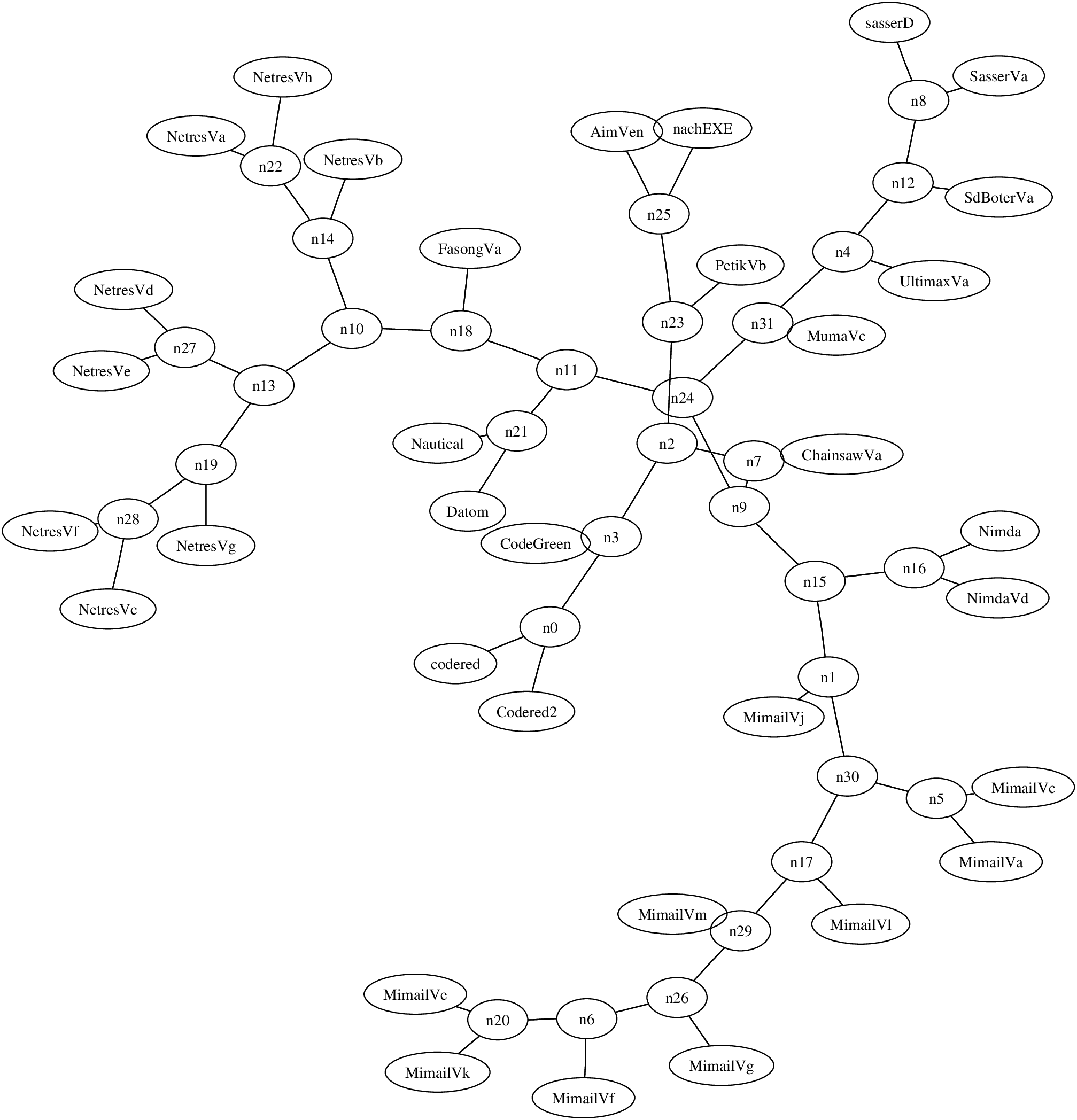, width=\columnwidth}
\caption{Windows Worms}
\label{windowsWorms}
\end{center}
\end{figure}

As text based worms are easy to analyze manually, we concentrate on binary executables.
We first of all clustered a variety of windows worms, which were not UPX compressed.
Surprisingly even binary worms cluster quite well, which indicates that they are indeed very similar.
The compressor seems to find a lot of similarities that are not immediately obvious by manual inspection.
Figure~\ref{windowsWorms} shows that the two available Sasser variants were clustered nicely together. The same goes for
the two versions of Nimda. This shows that classification by compression might be a useful approach.

\subsubsection{The effect of UPX compression}

Many worms found in the wild are compressed executables, where the compression is often modified to prevent
decompression. This makes it much harder to obtain the actual binary of the worm. But without this binary, we
cannot compare text strings embedded in the executable or disassemble the worm to apply an analysis of the program
flow. If a string is already compressed very well it is almost impossible to compress it much more. Thus we expected
initially that we would be unable to apply our compression based method with any success. Surprisingly, however, 
we are still able to cluster UPX compressed worms reasonably well. This indicates that UPX, which allows the binary to
remain executable, does not achieve the same strength of compression as bzip2.

\begin{figure}[p]
\begin{center}
\epsfig{file=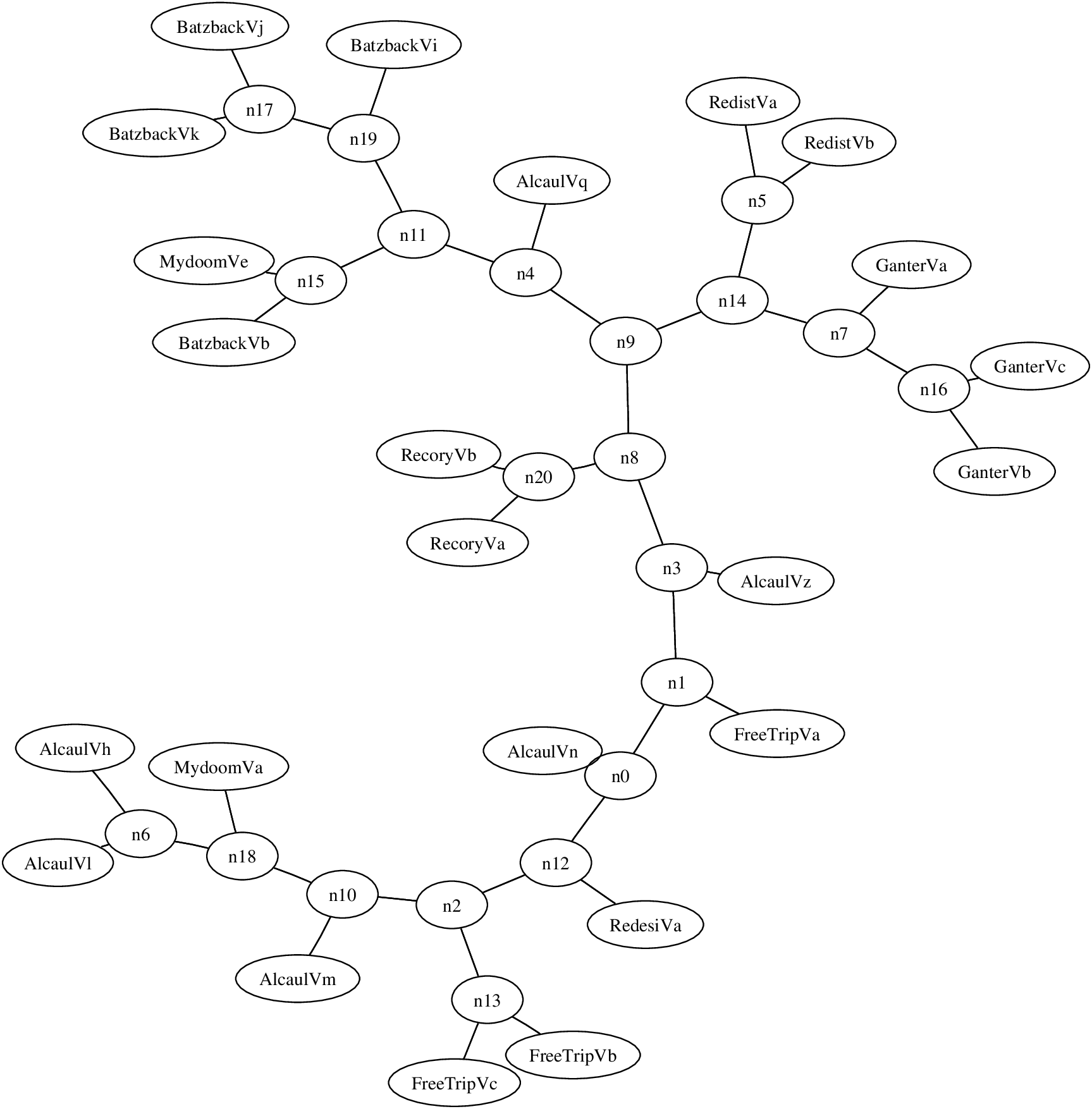, width=\columnwidth}
\caption{UPX Compressed Windows Worms}
\label{upxcompressed}
\end{center}
\end{figure}

\begin{figure}[p]
\begin{center}
\epsfig{file=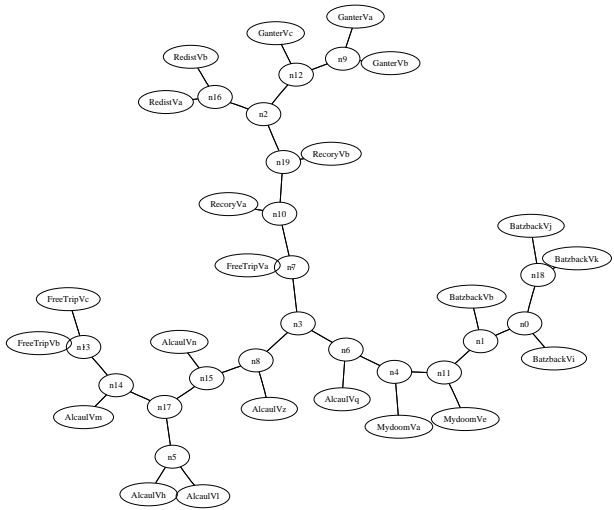, width=\columnwidth}
\caption{UPX Compressed Windows Worms after Decompression}
\label{decrypted}
\end{center}
\end{figure}

Figure~\ref{upxcompressed} shows a clustering experiment with worms which found UPX compressed initially. 
It shows that even though our method becomes less accurate, we still observe clustering, as for example 
in the case of the Batzback series. For other worms, such as Alcaul.q, clustering worked less well. This indicates
that even if worms are UPX compressed, we still stand a chance of identifying their family by simple compression.
Figure~\ref{decrypted} shows the result of the clustering process, after removing UPX compression. We now observe
better results.

However, it is not always possible to simply remove UPX compression before the analysis of the worm.
UPX compression can be modified in such a way that the worm still executes, but decompression fails.
This makes an analysis of the worm a lot more difficult.
We therefore investigated
how UPX compression scrambling influences the clustering process. All worms
in our next experiment
were selected to be UPX compressed and scrambled in the wild. Of the Alcaul
series, only Alcaul.r was scrambled. We included the other UPX compressed
Alcaul worms, to determine whether having part of the family UPX compressed
and scrambled and part of it just plain UPX compressed would defeat the
clustering process.

\begin{figure}[p]
\begin{center}
\epsfig{file=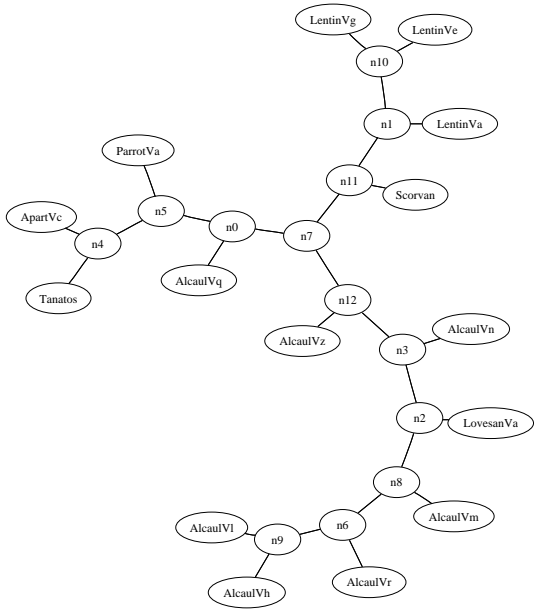, height=\columnwidth}
\caption{UPX Compressed and Scrambled Windows Worms}
\label{upxScramble}
\end{center}
\end{figure}

As Figure~\ref{upxScramble} shows, our method could be used to give a first indication to the family of the
worm, even if it cannot be UPX decompressed without great manual effort. Our Alcaul.r worm
which was scrambled clusters nicely with the unscrambled Alcaul worms. Also the Lentin series
is put together. 

\subsubsection{Worms and Legitimate Windows Programs}

\begin{figure*}[p]
\begin{center}
\epsfig{file=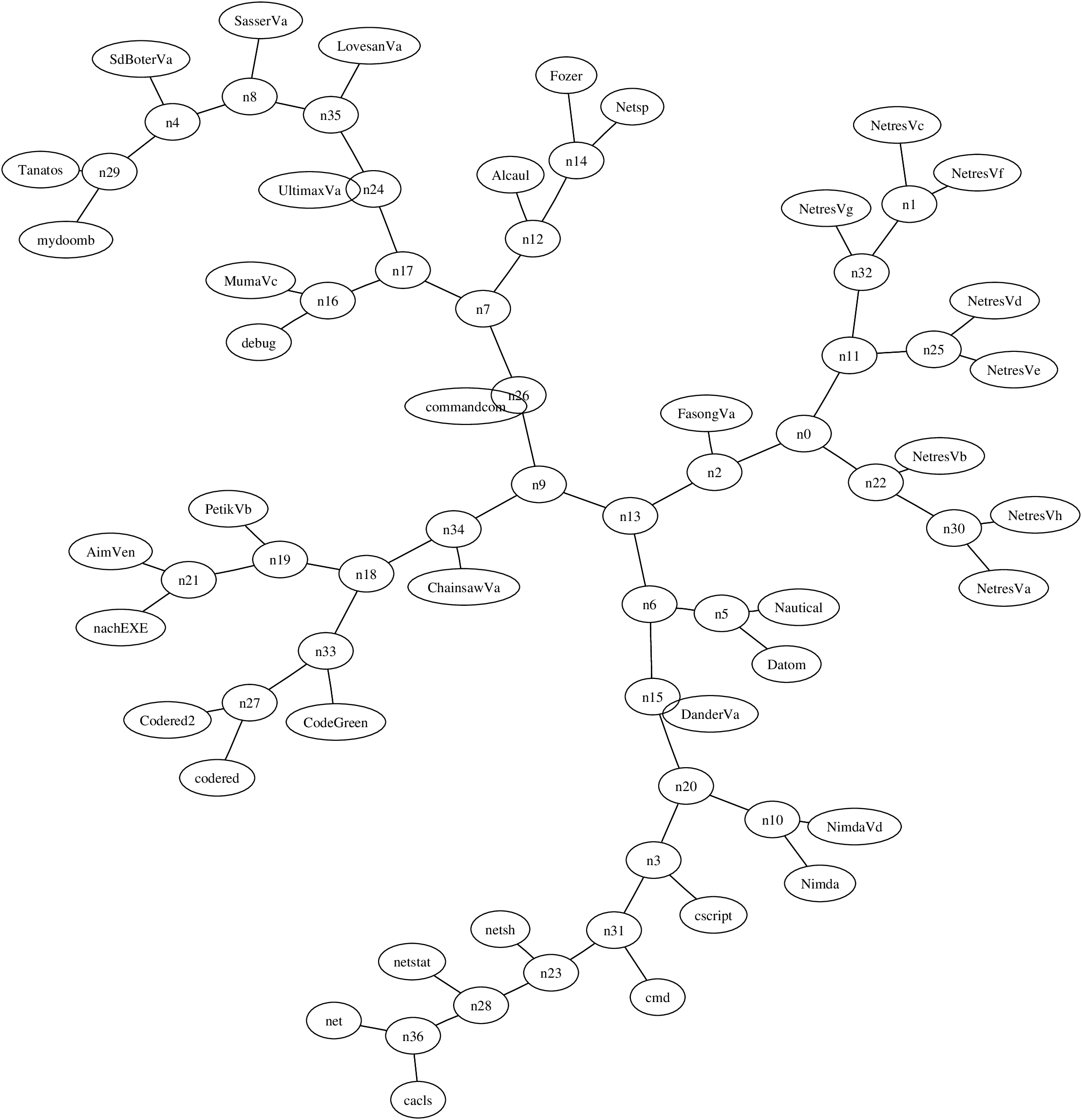, height=15cm}
\caption{Worms and Legitimate Windows Programs}
\label{legal}
\end{center}
\end{figure*}

Finally, we examine existing windows programs and their relation to internet worms.
For this we collected a number of small executables from a 
standard windows installation and clustered them together with various different types of worms. Figure~\ref{legal}
depicts the outcome of our test. The legal windows programs in use where: net, netsh, cacls, cmd, netstat, cscript,debug.exe and 
command.com. Surprisingly, our experiment shows most of the windows programs closely together. It even appears that
programs with similar functionality appear closely related. Only debug.exe and 
command.com are placed elsewhere in the tree, which we attribute to the fact that 
they have quite different functionality compared to the other programs. 
It would, however, be premature to conclude from this experiment that compression can always
be used to distinguish worms from legal windows programs. Our test included only a very limited
selection of programs. However, it seems to suggest that programs with similar functionality are
closely related. For example, CodeGreen is very close to CodeRed, which may be due to the fact
that it exploits the same vulnerability.

\subsection{Classifying Worms}\label{classifyWorms}

We are now trying to guess the family of an unknown worm.
Determining the family of a worm can make further analysis a lot easier. 
Again, we do not search for specifically selected patterns in the executables
or apply an analysis of the actual program. Instead, we simply use the NCD. 
To find the family of a worm, we compute the NCD between the worm and all
available binary worms. The best match $t$ is the worm which was closest to $w$ in terms of
NCD. We then set the family of $w$ to $t$'s family. In short, we assign the new worm $w$ to family $F$ such that
for $\exists t \in F, \forall k \in W, \mbox{NCD}(w,t) \leq \mbox{NCD}(w,k)$ where
$W$ is the set of all test worms. To avoid bad matches, we assign the label ``unknown'' to a worm
if $\mbox{NCD}(w,t) \geq 0.65$. This value was determined by experiment from our data set and
may need to be adjusted in a different situation.

In our first experiment we use 719 different worms, binary and non-binary. Of 284 of these worms, we had more than one representative
of each family available. For every worm, we determined its closest 
match among the remaining worms.
In 179 of the 284 cases were matching to its family would have been possible, we have succeeded. We obtained 39 bad matches.
This may seem to be a very low number of correct assignments. Note, however, that we here matched all types of worms against each 
other and used a very large set of worms which occurred in only one version. In practice, one could restrict the search for the family of a worm
to a more selected data set to obtain better results. For example, consider only worms which propagate via email.
Restricting our attention to worms labeled "I-Worm" alone, we can obtain a better classification as shown in Table~\ref{matchTab}. We used 454 of such worms, where
we had more than one representative of a family for 184 of these worms. Now only 13 worms
were classified incorrectly.

\begin{table*}[htbp]
\centering
\begin{tabular}{|c|c|c|}
\hline
Match & 454 Worms (184 in Family) & Average NCD \\
\hline
\hline
Good Family & 106 & 0.68\\
\hline
Bad Family & 13 & 0.43\\
\hline
No Family Found & 335 & 0.84\\
\hline
\end{tabular}
\caption{Classifying Worms}
\label{matchTab}
\end{table*}

Our experiment makes us quite hopeful that this simple method could be of practical use in the future.

\section{Analyzing Network Traffic}\label{traffic}

A lot of effort has been done to detect attackers on a network. Next to a large number of available intrusion detection
systems (IDS), intrusion detection is still a very active area of research. Intrusion detection can be split into two
categories: host based intrusion detection which focuses on detecting attackers on a certain machine itself and
network based intrusion detection which aims to detect attackers by observing network traffic. Many IDS use a combination
of both. In this section, however, we are only concerned with network based intrusion detection.

The most commonly used approach in existing systems is signature matching.
Once a known attack pattern has been identified, a signature is created for it. Later network
traffic is then searched for all possible signatures. 
A popular open source IDS system (Snort~\cite{snort}), for example,
offers this approach.
Clearly this works very well to identify
the exact nature of the attack. However, it first of all requires a lot of manual effort: prior 
identification of malicious data and signature creation becomes necessary. Secondly, if the number 
of signatures is large it is hard to check them all on a high volume link. In Section~\ref{complex}
we present a very simple compression based approach which does not suffer from these problems.
Finally, if a worm slightly alters its behavior it will no longer correspond to the signature and 
escape detection. In Section~\ref{snort2} we examine an approach which is able to recognize attack patterns
which are similar to existing ones.

Signature matching
is sometimes also referred to as misuse detection, as it looks for specific patterns of abuse. We speak of anomaly detection
if we are interested in identifying all kinds of anomalies, even caused by new and unknown forms of attacks.
Signature based schemes can also be used for anomaly detection. For example, an IDS has a profile
of typical behavior of each user. It then tries to match observed behavior to the profile. If the deviation
is too large an alarm is triggered.
Various efforts have also been made to apply neural networks for anomaly detection. Neural networks are
often used to perform off-line analysis of data, as they are computationally very expensive. This makes approaches based
on neural networks not very widespread so far.
In the following, we examine the use of compression as a tool to detect anomalies in network traffic.

\subsection{The Complexity of Network Traffic}\label{complex}

\subsubsection{The Complexity of Protocols}
The simplest approach we use is to estimate the complexity of different types
of traffic. We thereby try to answer the following question: Examining sessions on a specific port,
can we determine whether one kind of traffic has been replaced by another?
This may be useful to detect the success of certain remote exploits. For example,
we may only observe SSH traffic on port 22. After an attacker successfully exploited
the SSH server he may obtain a remote shell on the same port. His session is then
very different from a normal SSH session.

The method we use here is again extremely simple. For different protocols we identify the average compression ratio
of a session $S$: We take the payload of a session $S$ and compress it using
bzip2. The length of the compressed session $S_{compresssed}$ gives us an estimate for the complexity of
$S$, $C(S)$. We then calculate the compression ratio $R$ as
$$
R = \frac{\ell(S_{compressed})}{\ell(S)}.
$$
Intuitively, $R$ tells us how well $S$ compresses. If $R$ is small, $S$ compresses well and
thus has a low complexity. If $R$ is close to 1, $S$ can hardly be compressed at all and its
complexity is very high.

In order to apply this method in practice, the average compression ratio of good sessions on a certain port needs to be determined.
In order to find
anomalies in a new session, we extract the session from network traffic and compress it.  
If its compression ratio deviates significantly from the average compression ratio of expected traffic,
we conclude that an anomaly has occurred.

For the following comparison, traffic was collected on a small site providing web, email and shell
services to about 10 users and an ADSL uplink. The tag ``nc'' is used for terminal interactions.
The payload of sessions was extracted into separate files using chaosreader and compressed using bzip2.
We then calculated the compression ratio of each file. From this we computed the average compression
ratio and standard deviation of this ratio of all files belonging to a certain protocol.

\begin{figure}[htb]
\begin{center}
\epsfig{file=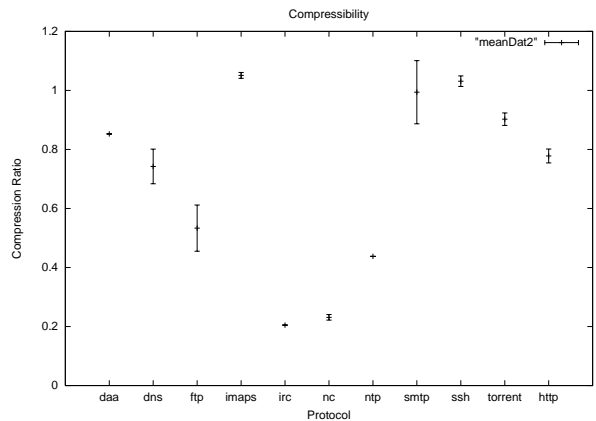, width=\columnwidth}
\caption{Average compressibility and standard deviation}
\end{center}
\end{figure}

As expected the purely text based protocols, such as IRC and nc compress very well. Protocol sessions
that carry encrypted data or binary data which is usually already compressed, on the other hand,
can not be compressed very well. For example, SSH and IMAPS sessions, carrying an encrypted payload, hardly
compress at all. Neither do HTTP and torrent sessions which carry compressed files.
Why does SMTP
traffic compress so badly? Closer inspection revealed that most captured SMTP sessions carried compressed
viruses instead of text messages. It may therefore be desirable to examine only the non-data part of
an SMTP session. We see that the standard deviation of SMTP traffic is quite high, which is caused by the
fact that emails containing compressed attachments compress muss less well than plain text messages.

This initial test shows that this method may well be applicable to separate protocols which
differ greatly in their complexity. For example, we could tell an IRC session from binary web
traffic. Likewise, we could identify a terminal interaction from the background of SSH traffic.
However, this method is unsuited for distinguishing protocols for file transfer or protocols
carrying encrypted data. We furthermore note that averages for web and SMTP traffic may vary
greatly depending on the web server accessed. Therefore averages would need to be determined for
every site individually. 
We also hope to use this method to detect new and unknown protocols an
attacker may use as long as their compression ratio
differs from the average compression ratio we expect. 

\subsubsection{Detecting complexity differences using Snort}\label{snort}

In order to test for complexity differences, we provide the compress plugin for the popular open-source IDS system 
Snort~\cite{snort}. We can use it to specify a lower and upper limit on the allowed compression ratio. 
If the observed traffic data falls outside this window, an alarm is triggered. The following rule
will trigger an alarm if the compressibility of traffic on port 22 drops below 0.9 and use zlib for compression.  

\smallskip
\begin{small}
alert tcp \$HOME\_NET any -$>$ any 22 (msg:"Low complexity on port 22"; compress:moreThan 0.9,zlib; 
	flow:to\_client,established; classtype:bad-unknown; sid:2565; rev:5;)
\end{small}
\smallskip

In this example no explicit upper limit is given, so the default value of 2 will be used. The compression ratio can become
larger than 1, if the input is completely incompressible and will actually be enlarged by the compressor.
The compress plugin takes the following comma delimited arguments:
\begin{itemize}
\item moreThan $<$$x$$>$: legal traffic has a compression ratio of more than $x$, where $x$ is a number between 0 and 1.
\item lessThan $<$$x$$>$: legal traffic has a compression ratio of less than $x$.
\item $<\mbox{compressor}>$: type of compressor to use. Current values are zlib for gzip style compression and rle for a simple
simulation of runlength encoding.
\end{itemize}

\subsection{Identifying similar attacks using Snort}\label{snort2}

In order to determine whether the observed network traffic is similar to an existing attack 
pattern, we again make use of the normalized compression distance (NCD) from Section~\ref{ncd}. 
If the NCD between a known attack pattern and the observed traffic is small, we conclude that
a similar attack is in progress.
This method is used by the NCD plugin. 
Since single packets can be very small, it makes
sense to use the NCD plugin in combination with stream4 TCP stream reassembly and only examine fully
reassembled TCP sessions. 

In the following simple example, we have recorded a session of an attack using an older vulnerability of 
Apache using PHP 3.0.15 on FreeBSD. This used to be widely available exploit with the name phploit.c~\cite{phploit}. This exploit
comes with different options, as to whether a shell should be opened on the vulnerable host on a different port
following the attack, or whether a shell should be executed directly. In our recorded session, we used
the option of binding a shell to another port. We then made the following entry in our Snort configuration:

\smallskip
\begin{small}
alert tcp \$HOME\_NET any -$>$ webserver 80 (msg:"Close to phploit attack."; flow:only\_stream; ncd:dist 0.8, file recordedSession; classtype:bad-unknown; sid:2566; rev:5;)
\end{small}
\smallskip

This entry tells Snort to trigger an alarm on all TCP sessions which have an NCD of less than 0.8 to the data contained 
in the file recordedSession. A new attack using the same options triggered this rule with an NCD of 0.608 to the recorded session. 
Executing the attack with the option to execute a shell directly also triggered this rule with an NCD of 0.638. Mirroring the
entire website on the test server did not result in any false positives. We have thus succeeded in recognizing the new, but slightly
different attack.
Whereas the above variation on the attack could have been easily recognized by selecting a different pattern to look
for initially, it nevertheless illustrates that the NCD may indeed be very useful in the detection of 
new attacks. Selected patterns are typically very short, which can make it much easier to escape detection by creating a small variation of
the attack.

The plugin currently makes use of the zlib library for compression. Other compression methods
would be possible.
It 
currently takes a single argument: dist $<$$x$$>$, where $x$ is maximum safe distance before 
the rule is triggered. 

\section{Conclusion and Open Questions}

We analyzed the binary executables of worms of different families and clustered them by family. Our 
analysis shows that many worms cluster well. Our method even performs reasonably well, if the worms are UPX compressed 
and the compression has been scrambled, which makes conventional analysis much harder. This indicates 
that our method may be a very useful tool in the initial analysis of new found worms in the future.
Many improvements are possible. Different compression methods may give better results.
Secondly, one could pre-classify worms into email viruses, internet worms, irc worms and vb scripts. 
In Section~\ref{classifyWorms}, we showed that restricting ourselves to worms labeled ``I-Worm''
already gave better results. Right now our aim was to provide a tool for the analysis of 
worms. It may be interesting see how this method can be incorporated in an automated virus scanner for email.

We showed that different types of traffic have different complexity. This might be a simple and
effective tool to distinguish between legal and illegal traffic on the same port provided the
complexity of the traffic differs largely. It does not require a priori knowledge of specific
patterns which occur only in specific types of traffic. The compress plugin for Snort allows the detection 
of potentially malicious packets and sessions which fall outside the specified compressibility window. 
It would be interesting to see how this method performs under high load. Our simple approach also
compresses traffic only once. A better performance may be achieved by compressing traffic in a
continuous fashion. Some initial experiments which did not make use of snort showed this to be
rather promising. Compression is an expensive operation, but so is pattern matching. Using efficient
and simulated compression may outperform matching a large pattern list. 

Using the normalized compression distance (NCD), we can detect anomalies similar to existing ones. The
NCD plugin for Snort can sound an alarm, when the observed session is too close to a given attack
session in terms of the NCD. This comparison is expensive. Nevertheless, we believe that our approach
could be useful to detect new variations of attacks initially. Once such a variation has been recognized, 
new patterns can be extracted to be used in conventional intrusion detection systems. Our method
seems promising and leaves room for further exploration:
In this paper,
we have only made use of standard compression programs. This also limits us to comparing packets
and sessions exceeding a certain size so that they can still be compressed to a smaller size by these programs.
Perhaps a compression program specifically targeted at compressing small packets would give better results.
It would also be possible to create a faster version of an existing compressor by merely simulating compression,
because we never want to decompress again. 

Finally, our method can be defeated by an attacker who intentionally alters the complexity of the 
traffic he generates. For example, he may interleave random with the actual data, to artificially 
increase the complexity of this session. Nevertheless, we believe that there are many scenarios 
where this is not possible or makes an attack considerably more difficult.
  
\section{Acknowledgments}
We thank Scott A. McIntyre and Alex Le Heux for valuable advise. Thanks also to the people who helped to collect worm data: 
Anthony Aykut, Lukas Hey, Jorrit Kronjee, Laurent Oudot, Steven de Rooij and Donnie Werner. The author acknowledges support from the
EU fifth framework project RESQ IST-2001-37559 and the 
NWO vici project 2004-2009.


\begin{thebibliography}{10}

\bibitem{carrera:malware}
E.~Carrera and G.~Erd\'elyi.
\newblock Digital genome mapping - advanced binary malware analysis.
\newblock In {\em Virus Bulletin Conference September 2004}, 2004.

\bibitem{complearn}
R.~Cilibrasi, A.~Cruz, Julio B., and S.~Wehner.
\newblock Complearn toolkit.
\newblock http://complearn.sourceforge.net/index.html.

\bibitem{halvar:personal}
T.~Dullien.
\newblock Personal communication.

\bibitem{evans}
S.~Evans and B.~Barnett.
\newblock Network security through conservation of complexity, 2002.
\newblock MILCOM 2002.

\bibitem{halvar:malware}
H.~Flake.
\newblock Structural comparison of executable objects.
\newblock In {\em DIMVA}, pages 161--173, 2004.

\bibitem{kulkarni:management}
A.~Kulkarni and S.~Bush.
\newblock Active network management and kolmogorov complexity, 2001.
\newblock OpenArch 2001, Anchorage Alaska.

\bibitem{kulkarni:ddos}
A.~Kulkarni, S.~Bush, and S.~Evans.
\newblock Detecting distributed denial-of-service attacks using kolmogorov
  complexity metrics, 2001.
\newblock GE CRD Technical Report 2001CRD176.

\bibitem{paul}
M.~Li and P.~Vitanyi.
\newblock {\em An Introduction to Kolmogorov Complexity and Its Applications
  (2nd Edition)}.
\newblock Springer Verlag, 1997.

\bibitem{phploit}
Security.is.
\newblock Apache/php exploit.
\newblock http://packetstormsecurity.org/0010-exploits/phploit.c.

\bibitem{tbaum}
A.~S. Tanenbaum.
\newblock {\em Computer Networks, 3rd edition}.
\newblock Prentice-Hall, 1996.

\bibitem{snort}
Snort~Development Team.
\newblock Snort. the open source network intrusion detection system.
\newblock http://www.snort.org.

\bibitem{virus}
Viruslist.com.
\newblock Virus encyclopedia.
\newblock http://www.viruslist.com/eng/viruslist.html.

\bibitem{rudi}
P.~Vitanyi and R.~Cilibrasi.
\newblock Clustering by compression, 2003.
\newblock http://arxiv.org/abs/cs.CV/0312044.

\end{thebibliography}
\end{document}